\documentstyle[12pt,aasms]{article}

\begin{document}

\title{
THE PHOTODISINTEGRATION OF COSMIC RAY NUCLEI BY SOLAR PHOTONS: THE
GERASIMOVA-ZATSEPIN EFFECT REVISITED
}

\author{Gustavo A. Medina-Tanco$^{1,2}$}

\begin{center}
and
\end{center}

\author{Alan A. Watson$^{2}$}

\affil{ 1.
Instituto Astron\^omico e Geof\'{\i}sico, University of S\~ao Paulo, Brasil
\\
gustavo@iagusp.usp.br
}

\affil{ 2. Dept. of Physics and Astronomy, University of Leeds,
Leeds LS2 9JT, UK
\\
a.a.watson@leeds.ac.uk
}

\singlespace

\begin{abstract}

The interesting possibility of measuring the masses of high energy
cosmic ray particles by observing pairs of extensive air
showers arriving at the earth nearly simultaneously was proposed
some years ago by Zatsepin (1951) and Gerasimova and Zatsepin (1960).
Such showers
would be created by the nuclear fragments originating as a result of
the photodisintegration of massive nuclei interacting with the solar
radiation field. In this paper we re-visit this possibility
in the context of existing and proposed detectors of high and
ultra-high energy cosmic rays considering a simple, yet realistic,
model of the interplanetary magnetic field.  The possibility of
observing the mass fragmentation of cosmic rays directly, however,
remains challenging.


\end{abstract}

\clearpage

\section{Introduction}

Cosmic ray nuclei travelling towards the Earth can interact with the
solar radiation field and photodisintegrate (Zatsepin, 1951), leading
to the almost simultaneous creation of pairs of extensive air showers
(EAS) in the earth's atmosphere.  If such pairs of showers could
be detected and the initiating energies measured then the ratio of the
greater energy to the lesser energy would give directly the mass of
the heavier fragment.  This assumes, as is most probable, that
photodisintegration produces a single nucleon as one of the
fragments.  Thus in principle the mass of incoming cosmic rays
could be measured rather directly.  Gerasimova and Zatsepin (1960)
studied this phenomenon in a
simplified analytical way which was
made necessary by the complexity of the problem and the limited
computing power then available.
As the problem of the mass composition of cosmic rays above the knee
of the spectrum remains as important and controversial an issue as it
was in the sixties (e.g., Watson 1997) it is of interest to assess the
potential of any mechanism that could help to tackle the problem. With
its remarkable simplicity, the Gerasimova-Zatsepin mechanism certainly
falls in this category.

Our main contribution is to address an important issue which was
treated incompletely in the initial work namely
the effect of the interplanetary magnetic field on the distribution
of the expected core separations.  We have made a significant
improvement in this regard by performing exact orbit integrations in
a realistic model of the interplanetary magnetic field.  We also use
modern estimates of the primary energy spectrum to calculate the
expected rate of arrival of such pairs of showers and discuss the
possibility of detection with current and planned instruments.

The relevant photonuclear interactions are discussed, for example, in Danos
and Fuller (1965), Hayward (1970) or Puget, Stecker and Bredekamp (1976).
The energy range of interest in the rest frame of the cosmic ray nucleus
spans from approximately 10 MeV up to 150 MeV. However, it is in the region
from $\sim 10$ MeV to $\sim 30$ MeV, the domain of the giant resonances
of the nuclear photoeffect, where most of the interactions result in the
emission of single nucleons, although two-nucleon emission can also
occur but with much reduced probability.

In the next section we describe our calculations and assumptions
and discuss the results, while our brief concluding remarks are
left to the last section.

\section{Numerical calculations and discussion of results}

In dealing with this problem, two different aspects must be considered:
the expected event rate and the relative deflection of the fragments.

In the original paper (Gerasimova and Zatsepin 1960) a deviation
of the order of 10$^{-2}$ cm is quoted for an assumed homogeneous
interplanetary field of the order of 10$^{-5}$ Gauss. This value is
in error by several orders of magnitude for a typical particle
with $\gamma \sim 10^{7}$ interacting with a photon in the vicinity
of the Sun.  An elementary calculation shows that deflections of
hundreds of kilometers should be expected at Earth for
a $\gamma \sim 10^{7}$ Fe nucleus losing a proton by
photodisintegration at 1 AU.  The oversight was subsequently
recognized by Zatsepin, as reported by Ginzburg and Syrovatsky (1964),
but no detailed calculations seem to have been made subsequently which
include
the effects of deflections in the interplanetary magnetic field.

For our calculations we use a realistic, yet simple, model of the
interplanetary magnetic field (IPMF) due to Akasofu, Gray and Lee
(1980). This model takes into account four different components to
describe the spiral 3-dimensional structure of the IPMF inside the
central 20 AU of the Solar System: (1) the dipole component
({\bf B$_{\mbox{dipole}}$}), (2) the sunspot component
({\bf B$_{\mbox{sunspot}}$}), (3) the dynamo component
({\bf B$_{\mbox{dynamo}}$}) and the ring current component
({\bf B$_{\mbox{ring }}$}). The total field is thus expressed as:

\begin{equation}
\vec{B} = \vec{B}_{\mbox{dipole}} + \vec{B}_{\mbox{sunspot}} +
          \vec{B}_{\mbox{dynamo}} + \vec{B}_{\mbox{ring }}
\end{equation}

The solar dipole moment is well known ($\sim 3.4 \times 10^{-32}$
Gauss cm$^{3}$). The dynamo component originates in a poloidal
current system which exits the Sun at its poles,
reaches the heliosphere at high latitudes, flows over this
surface towards the ecliptic plane and finally closes the circuit
through an inward equatorial current.  This system is supposed to be
generated by a dynamo process induced by the solar rotation in the
dipolar field.  The ring current component is produced by a thin
equatorial sheet current that extends up to the heliopause. The
sunspot component constitutes the magnetic arcade immediately above
the photosphere.  In the model of Akasofu, Gray and Lee (1980) this
component is represented by an ensemble of spherical dipoles just
below the surface. Its main purpose is to allow the connection of all
the field lines of the equatorial sheet to the Sun's surface without
significantly distorting the solar dipole field.
Both the solar dipolar component and the sunspot component decrease
as $\propto r^{-3}$ and therefore, outside the coronal region,
the IPMF is mainly determined by the dynamo and ring current
contributions. In figure 1a we show the magnitudes of the
total IPMF, and of the ring and dynamo components as a function
of the cylindrical coordinate $\rho$ over a plane located at
z=0.01 AU above the ecliptic.  Figure 1b shows the corresponding
cylindrical components of the IPMF. The signs over the curves
indicate the sign of each component. Note, however, that all the
components of the magnetic field reverse their direction with the
11 year solar cycle.

The interplanetary medium is permeated by this magnetic field
and by the photon radiation field originating at the Sun's
photosphere.  For our purpose it is sufficient to assume that the
radiation field spectrum is that of a black body at
$T_{eff} = 5770$ K.  Consequently, the number density of photons
at a distance $r_{AU}$ (in astronomical units) from the Sun is:

\begin{equation}
n(\epsilon_{eV}) d\epsilon_{eV}
         \sim 7.8 \times 10^{7}
                  \times \frac{1}{r_{AU}^{2}}
                  \times \frac{\epsilon_{eV}^{2} d\epsilon_{eV}}
                              {exp(\epsilon_{eV}/0.5)-1}
         \; \; \; \; [\mbox{cm}^{-3}]
\end{equation}

\noindent
where $\epsilon_{eV}$ is the energy of the photons in eV in the
reference system of the Sun.

Every cosmic ray particle penetrating the Solar System must
traverse this environment before reaching the Earth.
In the rest frame of the nucleus the energy of the photons is
boosted to:

\begin{equation}
\epsilon_{N} = \epsilon \left( \gamma + \sqrt{\gamma^{2}-1}
                               \cos \alpha \right)
             \sim 2 \gamma \epsilon \cos^{2} \frac{\alpha}{2}
\end{equation}

\noindent
where $\alpha$ is the angle between the momenta of photon and nucleus
in the Sun's reference system (e.g. Gerasimova and Zatsepin 1960).
When the energy of the photons in the frame of the nuclei is larger
than some few MeV, the nuclei can undergo photodisintegration.
This process is most important between $15$ and $25$ MeV, in the
region of the peak of the giant dipole resonance, although there is
still a significant contribution to the cross section from energies
beyond $25$ MeV up to the threshold for photo-pion production at $\sim
145$ MeV (Puget, Stecker and Bredekamp 1976, Hillas, 1975). In the
lower energy band, i.e. in the region of the giant dipole resonance
(say, $\epsilon \widetilde{<} 30$ MeV), mainly single nucleons are
emitted although double nucleon emission also takes place. At higher
energies non-resonance processes are responsible for multinucleon
emission.  The corresponding cross-section parameters and branching
ratios can be found in Puget, Stecker and Bredekamp (1976).
Gerasimova and Zatsepin (1960) use a form for the cross-section that
includes only the giant dipole resonance. We use the Gerasimova
and Zatsepin representation of the cross-section at low energies, but
have included additionally the non-resonant contribution at higher
energies up to the photo-pion production threshold.  In Figure 2 we
compare the cross-sections for Fe as calculated by Gerasimova
an Zatsepin (1960), Puget, Stecker and Bredekamp (1976), an
approximation by Hillas (1975), and the cross section used in the
present work.  It is evident that, although the Gerasimova and
Zatsepin cross section is wider than the Lorentzian function used in
Hillas's approximation, it compares rather well with the combination
of single and double nucleon emission as given by Puget and co-
workers.  Our approximation should give an upper limit to the
photodisintegration rate.  Uncertainties in the photodisintegration
cross section will translate to only some few 10\% and this does not
alter our conclusion significantly.

The emission of the nucleon(s) can be assumed to be isotropic in the
reference system of the nucleus.  Transforming to the Earth
reference system, the emission of the fragments is concentrated
within a cone of aperture $\sim 1/\gamma$ around the original
direction of propagation of the parent nucleus.  Therefore, at
the high Lorentz factors of interest here ($\gamma > 10^{7}$),
we can assume that both fragments have, after the interaction
with the photon, exactly the same direction as the incoming
nucleus.  Hence, we calculate relative deflections solely as the
product of the differential bending of the fragments due to the
action of the IPMF.  This is, in fact, opposite to the approach
taken by Gerasimova and Zatsepin.  They neglected the effect
of the IPMF and assumed that the distribution function of shower
core separations was given only by the angular distribution of the
fragments Lorentz-transformed to the Earth rest frame.

We consider a spherical volume surrounding the Earth and
extending up to $r_{max} = 4$ AU.  The photon density is too low and
the fragment deflections too large for any significant contribution to come
from outside this region. A grid is constructed giving the
fragment separation at Earth for a parent Fe nucleus interacting at any
point inside the volume.  As we are more interested in an upper limit
than in an accurate calculation, we further assume that all of the
cross-section goes into single nucleon emission producing, as
daughters, both a Mn nucleus and a proton.

To present the results we have chosen a polar coordinate system
analogous to the galactic coordinate system. The reference system is
centered on the Earth and its equatorial plane coincides with the
ecliptic plane.  The Sun is located at the origin of both
longitude $\phi$ and latitude $\theta$.  Latitudes are positive to
the North, while longitudes are positive to dusk and negative to
dawn.  Distances, in AU, are measured outwards from the Earth.

In figure 3 we show core separation distribution functions for
three particular directions on the sky: noon ($\phi \sim 1.3$ deg
-- i.e., a $5$ R$_{\odot}$ perihelion),
midnight ($\phi \sim -180$ deg), and mid-afternoon ($\phi \sim 45$
deg). For these plots the incident primaries are Fe nuclei of total
energy $E = 6.3 \times 10^{17}$ eV.  This energy was chosen to be
near the maximum of the fragmentation cross section for
interaction with $\epsilon \sim 1$ eV photons (a typical energy
of the solar radiation field photon).  The three curves are
proportional to the number of Gerasimova-Zatsepin events
coming from each direction, and have been normalized such that
the frequency is 1 for the smaller separation $\delta$ arriving
from the noon-side.  It can be seen that the effect of the IPMF is
much larger than that arising from the transverse separation of the
fragments leaving the interaction: at 1 AU the angular spread of the
fragments gives a separation of about 15 km.  In fact
the separation produced by the magnetic field is so large that
the possibility of observing both partners of a disintegration
process is rather limited at $\gamma \sim 10^{7}$ for any existing
detector.

The mean free path of a nucleus against photodisintegration is
given by:

\begin{equation}
\frac{1}{\lambda(l)} = \int_{0}^{\infty}
            n(l,\epsilon)
            \times
            \sigma_{frg}
            \left\{
                  2 \gamma \epsilon \cos^{2} \left[
                                             \frac{\alpha(l)}{2}
                                             \right]
            \right\}
            \times
            2 \cos^{2} \left[
                       \frac{\alpha(l)}{2}
                       \right]
            d\epsilon
\end{equation}

\noindent
where $l$ is the coordinate along the path of the nucleus, $\epsilon$
is the photon energy in the rest frame of the Sun, $\sigma_{frg}$ is the
fragmentation cross-section specified previously, and $\alpha$ is
the angle between the propagation directions of the nucleous and
the photon, the latter being taken as a heliocentric radiovector.

Therefore, defining $\Phi_{\infty}$ as the unperturbed incoming cosmic
ray flux at the external border, $r_{max} = 4 AU$, and $\Phi_{GZ}$
as the flux of Gerasimova and Zatsepin fragment pairs,
the relative {\bf GZ} flux is:

\begin{equation}
\eta_{GZ} = \frac{\Phi_{GZ}}{\Phi_{\infty}} = 1- \exp\
                                     \left[
                                      - \int_{r_{max}}^{r_{min}}
                                      \frac{dl}{\lambda(l)}
                                     \right]
\end{equation}

\noindent
where $r_{min} \sim 0.02$ AU is adopted as the inner spherical surface
up to which the integration is carried out.

Figures 4a and 4b show all sky maps of the ratio
$\eta_{GZ} = \Phi_{GZ} / \Phi_{\infty}$, i.e., of the fraction of
Gerasimova-Zatsepin events among the incoming cosmic ray flux.
The Sun is at the center of each figure, and the ecliptic plane
runs horizontally through the middle of the figure ($\theta = 0$).
The shaded function in the background is $\eta_{GZ}$, while the
contour lines indicate the medians of the separation between
the cores of correlated showers for each direction in the sky.
Note that the labels on the separation contours of figure 4a are
logarithmic, while those in figure 4b are linear. Both figures
are calculated for monoenergetic Fe nuclei at
$E = 6.3 \times 10^{17}$ eV (roughly the maximum of the
photodisintegration cross section). In Figure 4a no account has been
taken of any acceptance effects such as would be imposed in practice
by a detector system: every event is counted irrespective of the
separation of the showers.

    In Figure 4b we show only those shower pairs for which the
cores are separated by $\delta < 10$ km.  Such a separation is
relevant for the AGASA array (Chiba et al. 1992)
which has a collecting area of $100$ km$^{2}$.  From these two
figures it can be clearly seen that high values of $\eta_{GZ}$
only be obtained in the day-side in the vicinity of the Sun.
Furthermore, large values of $\eta_{GZ}$ arise only when no cut-off
in separation is considered: for any real instrument of finite
size (figure 4b) the flux is several orders of magnitude smaller.
In fact, so severe is this effect that when the separation is taken
into account( figure 4b), the maximum $\eta_{GZ}$ is obtained on the
night side of the earth around midnight. In other words, the small
number of photons nightward of the earth is more than compensated by
the high deflection on the day side which limits the effective
integration volume to regions very near the Earth. The decrease of the
effective integration volume on the day side can easily be seen by the
change in symmetry of the function $\eta_{GZ}$ in going from figure
4a to 4b. In the first case there is axial symmetry around the Sun
because most of the events originates in its vicinity. In figure 4b,
when only small $\delta$ events are accepted, they originate very near
the Earth, and therefore there is symmetry with respect to the
ecliptic plane, revealing the planar spiral topology of the
IPMF in the neighborhood of the Earth.

The low values of $\eta$ imply very low GZ fluxes.  To make an
estimate we assume that all nuclei above $E \sim 6 \times 10^{17}$ eV
are of iron and that the integral flux is $3.8 \times 10^{-12}$
$m^{-2}s^{-1}sr^{-1}$( a
value based on Fly's Eye and Haverah Park data).  Adopting an average
value of $\eta_{GZ}$ =$\ 10^{-6}$ (figure 4b) we find the rate of GZ
events to be $\sim 0.01$ per year on \newline 100 $km^2$ (the AGASA area).
Even for the detectors of the Auger Observatory (3000 $km^2$ per
site) only $\sim 0.3$ events per year with less than 10 km separation
would land on the array.  These rates are much too small for
detection in the case of AGASA and for the Auger observatory the events will
be
difficult to identify within the background as the shower from the
nucleonic fragment is likely to trigger only one detector because of
the 1.5 km separation planned for the array.  Our estimate of the rate is
several orders of magnitude less than that made in the original
examination of this effect in part because of the magnetic deflections
but also because modern estimates of the intensity at $\ 6\times
10^{17}$ eV are about 15 lower that believed in the late 1950s.

   The situation does not get better at higher energies even for the
planned large area detectors such as the Auger project and the
proposed OWL satellites. The latter experiment involves a pair of twin
satellites observing extensive showers in the upper atmosphere from
the outer space. The advantage of such an experiment is the huge
exposure area, $10^{5}-10^{6}$ km$^{2}$.  The disadvantage from the
point of view of GZ events is, however, that OWL observes the night
side of the atmosphere where the expected $\eta_{GZ}$ is much smaller.
Figures 5a and 5b illustrate the situation at higher energies: $E = 3
\times 10^{18}$ eV and $E = 3 \times 10^{19}$ eV respectively. In both
cases a maximum separation $\delta max = 1000$ km is used,
representative of a $10^{6}$ km$^{2}$ experiment like OWL. The medians
of the separations are conveniently smaller, but the GZ fluxes are
again too low to be of practical use.  Adopting an integral intensity
of  $2 \times 10^{-14}$ $ m^{-2}s^{-1}sr^{-1}$ and $\eta_{GZ}$ = $10^{-7}$
gives only 0.06
events per year for an area on $10^{6} km^{2}$.  Hence the GZ flux is
too low to make detection practical.  Furthermore, at very high
energies the separation between the cores is so small ($\delta \sim 5$
km on the night side at $3 \times 10^{19}$ eV, or even $\delta \sim
0.5$ km $\sim 10^{20}$ eV) that confusion might arise between the
signals associated with showers.

\section{Conclusions}

We have re-analyzed the proposal of Gerasimova and Zatsepin (1960)
of using pairs of correlated showers, originating in the photodisintegration
of heavy nuclei interacting with the solar radiation field, as a
mass-spectrometric technique.

We consider a simple but realistic model of the interplanetary magnetic
field to demonstrate that the magnetic deflection dominates the
distribution function of core separations.  The interactions with the
solar radiation field inside a sphere of 4 AU around the Earth are
calculated for incoming Fe nuclei under the simplifying assumption of
single nucleon emission. The results are presented as all-sky maps,
highlighting the considerable anisotropy of the solution.

From our calculations it is apparent that the events arising from
this very beautiful idea are too infrequent to be of use in any
real experiment, either in operation or currently proposed, as a mass
measuring technique.


This work was done with the partial support of the Brazilian agency FAPESP.

\newpage

\noindent
{\bf Figure Captions}

\bigskip

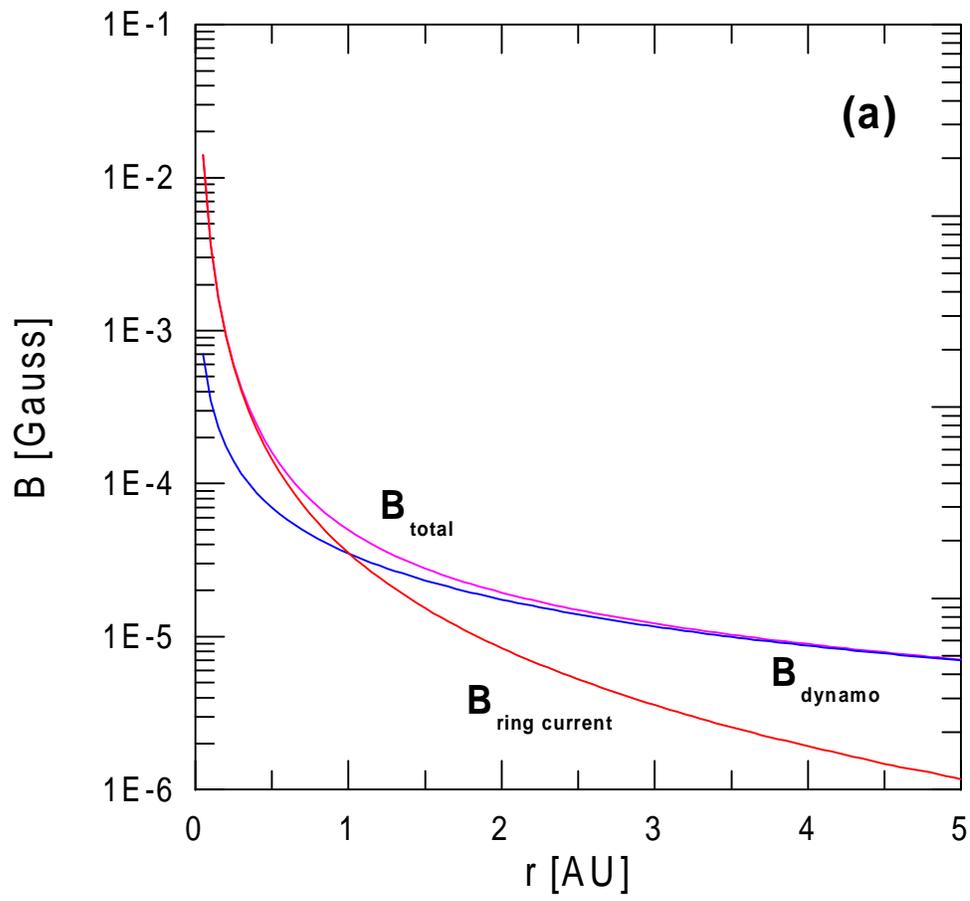
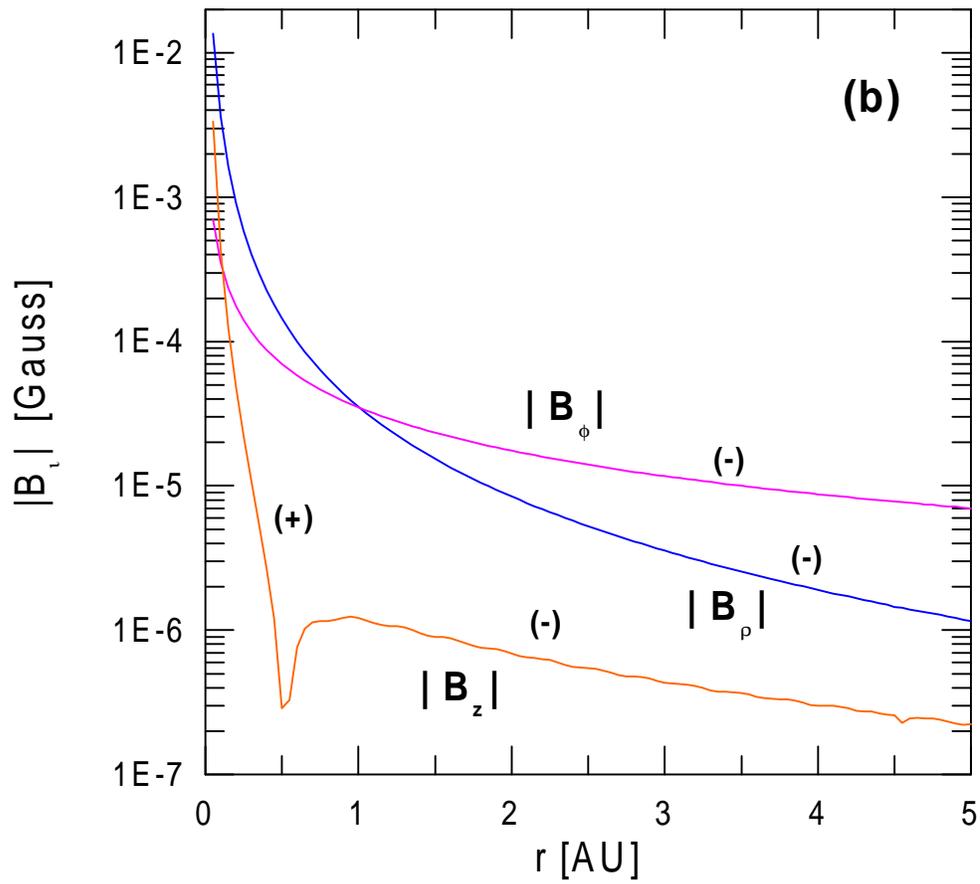
{\bf Figure 1:} (a) Total magnitude of the interplanetary magnetic field
and dynamo and ring current components as function of heliocentric distance
over a plane parallel to the ecliptic plane and located at z=0.01 AU. (b)
absolute value of the cylindrical components of the total magnetic field;
the signs of the several components are indicated above the corresponding
curves.


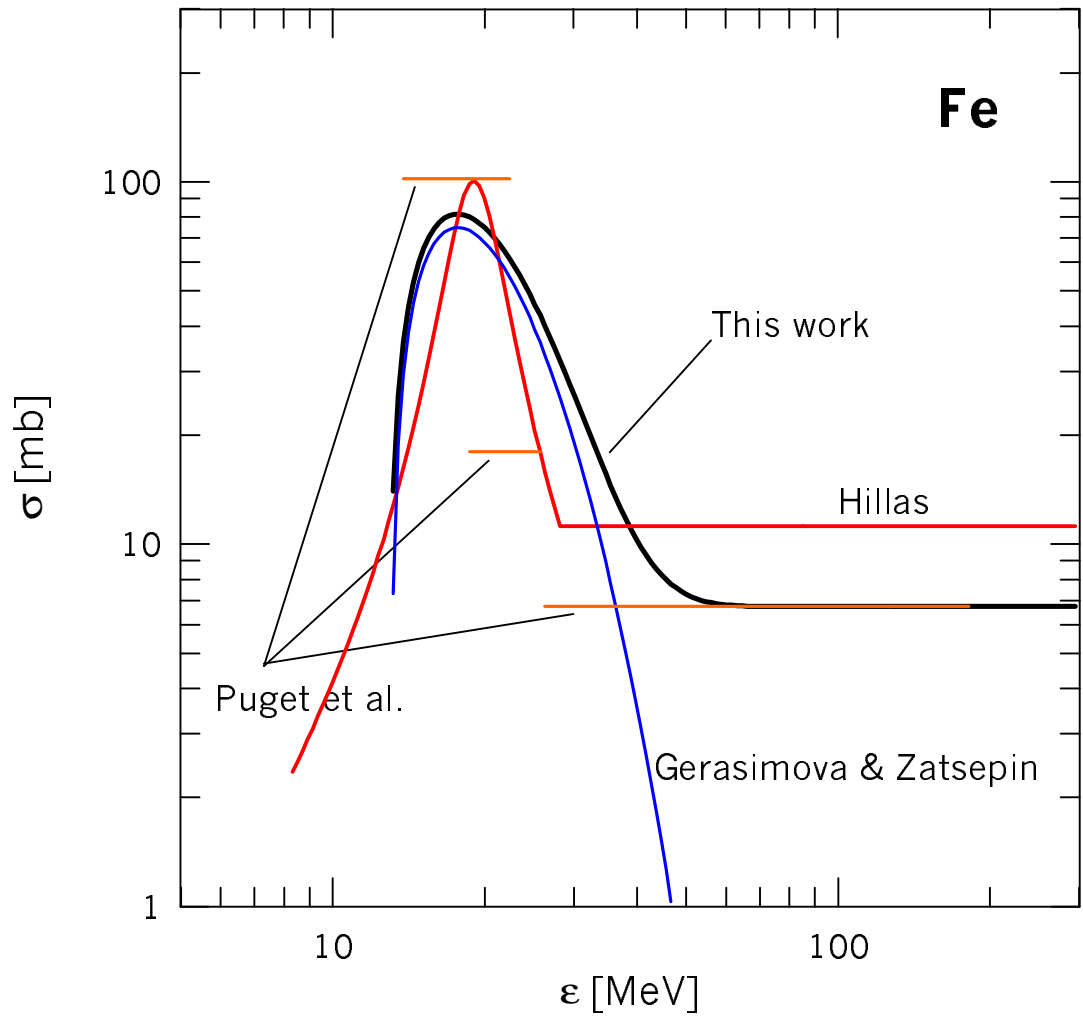
{\bf Figure 2:} Iron nuclei photodisintegration cross sections. See text
for details.


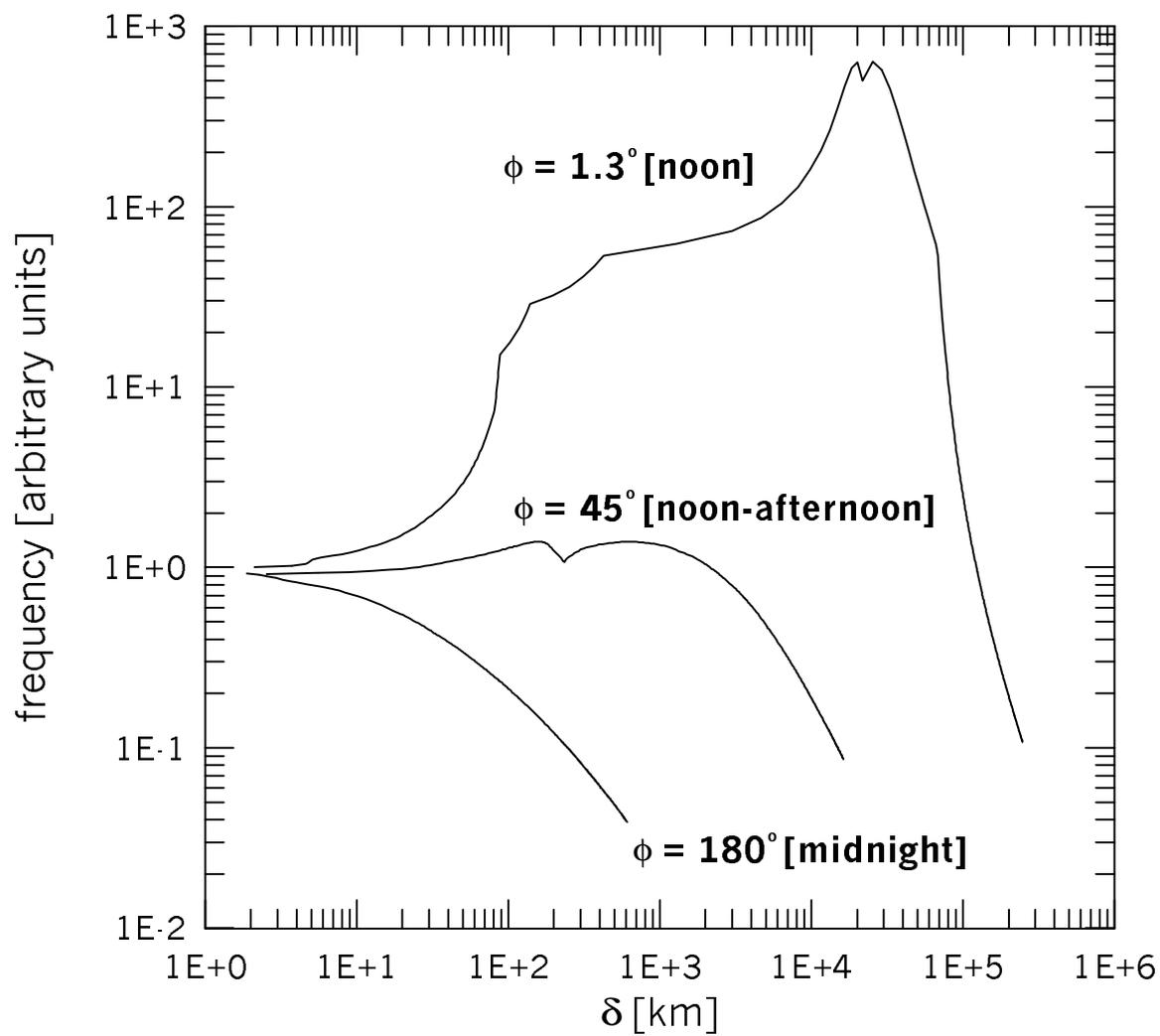
{\bf Figure 3:} Core separation distribution functions. All the curves
have the same normalization and can therefore be compared directly.
each curve corresponds to a particular direction in the sky, as
indicated in the respective labels, inside a solid angle of
$\sim 2 \times 10^{-3}$ sr.


{\bf Figure 4:} Fraction of GZ events, $\eta_{GZ}$, for Fe nuclei at
$E = 6.3 \times 10^{17}$ eV (a) regardless of the separation $\delta$
between correlated EAS and (b) only for $\delta < 10$ km. The contour
lines indicate the median of the separation $\delta$, and are labeled
logarithmically (log$_{10}$) in (a) and linearly in (b). The Sun is at the
center
of the image, while midnight is at the left and right borders of the
figures.


{\bf Figure 5:} As in figure 4, but for (a) $E = 3 \times 10^{18}$ eV
and (b) $E = 3 \times 10^{19}$ eV. The maximum separation allowed
is in both cases $\delta_{max} = 10^{3}$ km, and so this corresponds
to a hypothetical experiment with an effective area of $\sim 10^{6}$
km$^{2}$. The contour lines are labeled linearly with the median of
$\delta$ inside each solid angle.



\end{document}